\documentclass{article}
\usepackage{amssymb}
\usepackage{amsmath}

\begin{document}

AN ANALYTIC STUDY OF THE GRAVITATIONAL WAVE PULSAR SIGNAL WITH
SPIN DOWN EFFECTS\\

S. R. Valluri, Department of Physics \& Astronomy and Department
of Applied Mathematics, University of Western Ontario, London,
Canada, e-mail: valluri@uwo.ca

F. A. Chishtie, Department of Applied Mathematics, University of
Western Ontario, e-mail: fchishti@uwo.ca

Adam Vajda, Department of Physics \& Astronomy and Department of \
Computer Science, e-mail: avajda@uwo.ca

\begin{abstract}
In this work, we present the analytic treatment of the Fourier
Transform (FT) of the Gravitational Wave (GW) signal from a pulsar
including spin down corrections in a parametrized model discussed
by Brady et. al. \cite{BCCS98}. The formalism lends itself to a
development of the FT in terms of well known special functions and
integrals defining the spin down moments.
\end{abstract}

\section{Introduction}

The detection of gravitational waves (GW) from astrophysical
sources is one of the most outstanding problems in experimental
gravitation today. Large laser interferometric gravitational wave
detectors like the LIGO, VIRGO, LISA, TAMA 300, GEO 600 and AIGO
are potentially opening a new window for the study of a vast and
rich variety of nonlinear curvature phenomena. This network of
gravitational wave detectors can confirm that GW exist and
monitoring these wave forms give important information on their
amplitudes, frequencies and other important physical parameters.

The detection of GW necessitates the substantial accumulation of
Signal to Noise (S/N) over long observation periods. The data
analysis for continuous GW like, for example, from rapidly
spinning neutron stars is an important problem for ground based
interferometric detectors that demands analytic, computational and
experimental ingenuity.

In recent works \cite{JVD96,CQG2002} we have implemented the
Fourier transform (FT) of the Doppler shifted GW signal from a
pulsar with the Plane Wave Expansion in Spherical Harmonics
(PWESH). Spherical-harmonic multipole expansions are used
throughout theoretical physics. Indeed, they arise wherever one
deals with fields, be they electromagnetic, gravitational,
hydrodynamical and solid body etc. The expansion of a plane wave
in spherical harmonics has a variety of applications not only in
quantum mechanics and electromagnetic theory \cite{MWIEEE}, but
also in many other fields. In linear theories, such as vacuum
electromagnetic-wave theory, the problem is relatively simple: the
field's multipole components evolve independently of each other;
there is no coupling. However, in nonlinear theories like general
relativity, it is more difficult. A number of researchers have
used spherical-harmonic expansions for a variety of problems in
general relativity, including problems where nonlinearity shows up
as Kip Thorne \cite{KThorne80}has pointed out in detail. The
relationships between scalar spherical harmonics incorporated in
our present work and the various other harmonics like the pure
spin vector harmonics related to the Regge-Wheeler harmonics have
been discussed \cite{KThorne80}. The basis states in the PWESH
expansion form a complete set and facilitate such a study. It also
turns out that the consequent analysis of the Fourier Transform
(FT) of the GW signal from a pulsar has a very interesting and
convenient development in terms of the resulting spherical Bessel,
generalized hypergeometric function, the Gamma functions and the
Legendre functions. Significant analytic and numerical studies
have been carried out by many, for example, by Olver \cite{OLV},
R.C. Thorne \cite{RThorne}, van der Laan and Temme\cite{LT} and
Cherry \cite{CHERRY}. Hypergeometric functions also arise in the
analysis of gravitationally radiating binary stars \cite{PIERRO}.
The generalized hypergeometric functions in our analysis find
their extensions to the H-functions, Meijer G-functions \cite{DAV,
MS} and the Heun \cite{WHIT, SPEC} functions.

Both rotational and orbital motions of the Earth and spindown of
the pulsar can be considered in this analysis which happens to
have a nice analytic representation for the GW signal in terms of
the special functions above. The signal can then be studied as a
function of a variety of different parameters associated with both
the GW pulsar signal as well as the orbital and rotational
parameters. Regardless of the immediate usefulness of such an
analysis for GW data community which we do indeed look forward to
the numerical analysis of this analytical expression for the
signal offers a challenge for efficient and fast numerical and
parallel computation. The plane wave expansion approach was also
used by Bruce Allen and Adrian C. Ottewill \cite{AO96} in their
study of the correlation of GW signals from ground-based GW
detectors. They use the correlation to search for anisotropies
from stochastic background in terms of the $l, m$ multipole
moments. Our PWESH formalism enables a similar study. Recent
studies of the Cosmic Microwave Background Explorer have raised
the interesting question of the study of very large multipole
moments with angular momentum $l$ and its projection $m$ going up
to very large values of $l\sim1000$. Such problems warrant an
intensive analytic study supplemented by numerical and parallel
computation.

However, the important spin down corrections were not included in
our previous works. A considerable amount of work has been done on
the search for the continuous GW signals. Livas, Jones and
Niebauer \cite{LIV95} have investigated, in separate works, the
time series of continuous GW signals that incorporated modulation
of the motion of the detector over limited regions of parameter
space. They have not considered pulsar spin down, and restricted
their analysis to small areas of the sky. In this work, we present
a parametrized model clearly discussed by Brady et al
\cite{BCCS98,BC2000} for the gravitational wave frequency and the
phase measured at the ground based detector that includes the
spindown parameters.

Spin down corrections are important in the analysis of GW signals.
Brady et al \cite{BCCS98,BC2000} indicate that most of the S/N is
accumulated during the final stages of spin down of pulsars. There
could exist a class of pulsars which spin down mainly due to
gravitational radiation reaction \cite{BL}. The frequency scales
as $f \propto \tau^{-1/4}$ in these types of pulsars, where $\tau$
is the age of the pulsar. Assuming that the mean birth rate for
such pulsars in our galaxy is $\tau_{B}^{-1}$, the nearest one
should be a distance $r=\sqrt{\tau_{B}/\tau}$ from earth, where
$R\simeq10$ kpc is the radius of the galaxy. This estimate has
been provided by Brady et.al \cite{BCCS98}. We use the expression
for $h_c$ provided by Thorne \cite{KThorne87} and
\cite{JKS1998,LIGOR,SS2002}.

The amplitude, $h_c$ is given as,

\begin{equation}
h_c=\frac{16\pi^2G}{c^4}\frac{\epsilon I f^2}{r}
\end {equation}

where $G$ is the gravitational constant, $f$ is the sum of the
frequency of rotation of the star and the precession frequency,
$I$ is the moment of inertia with respect to the rotation axis,
$\epsilon$ is the poloidal ellipticity of the star and $r$ is the
distance to the star. Hughes et. al, suggest that $h_c
\leq10^{-24}$.

The frequency domain characteristic of the GW signal consists of
two components with carrier frequencies $f_0$ and $2f_0$
(Jaranowski et. al.\cite{JKS1998}) that are both amplitude and
phase modulated. The amplitude modulation in their analysis is
determined by functions which split each of the two components
into five lines that connect the GW and the earth rotation
frequencies. They observe that the frequency modulation (FM)
broadens the lines. They make estimates for a kilohertz GW
frequency, the spin-down age $\tau=40$ years, and an observation
time of 120 days for the maximum frequency shifts due to the
neutron star spin-down, Earth's orbital motion and Earth's diurnal
motion which are, respectively, of the order $\sim$ 8, $\sim$ 0.1,
and $\sim$ $10^{-3}$ Hz. In this paper, we have considered a
single component GW frequency, $f_0$ and neglected the amplitude
modulation.

The paper is organized as follows. In Section 2 we outline the FT
of the GW signal from the pulsars which is relevant to the
detection of continuous gravitational waves. We focus attention on
spin down effects which indicate significant frequency evolution
over periods of several weeks of observation. In this section, we
use the expression for a parametrized model of the expected
gravitational waveform, including modulating effects due to
detector motion. We derive the analytic form of the FT with spin
down corrections in the following section. The expressions for the
spin down moments are computed using the familiar technique of
differentiation with respect to a parameter. We present our
conclusions in section 4.

\section{Frequency and phase evolution with Spin Down}

Manchester (1992) and Kulkarni (1992) have suggested that pulsars
lose rotational energy by electromagnetic braking, particle
emission and emission of GW \cite{M92, KUL92}. Therefore, the
rotational frequency is not completely stable, and varies over a
timescale $t$ which is of order the age of the pulsar. Younger
pulsars with periods of tens of milliseconds have the largest spin
down rates. Current observations suggest that spindown is
primarily due to electromagnetic braking. Brady et. al. have
suggested a sufficiently general model of the frequency evolution
to cover all possibilities in the study of pulsar GW signal
detection. For observing times $t_{obs}$ much less than $t$, the
frequency drift is small.  We use a slightly modified form of the
parametrization given by Brady and Creighton (2000)\cite{BC2000}
in that it differs by an factor of $1/2$, and the summation index,
$k$ starts from zero. Our expression is the Doppler modulated
version of the earlier power series of the frequency evolution
given by \cite{BCCS98}. The parametrized model for frequency and
phase evolution with spin down corrections is given in the
following form.

\begin{equation}
f(t;\lambda)=\frac{f_0}{2}\left(1+\frac{\vec{v}}{c}\cdot\hat{n}\right)\left(1+\underset{k
= 0} \sum{f_k}
\left[t+\frac{\vec{r}}{c}\cdot\hat{n}\right]^{k}\right)
\end{equation}

\begin{equation}
\phi(t;\lambda)=\pi
f_0\left(t+\frac{\vec{r}}{c}\cdot\hat{n}+\underset{k =
0}\sum{\frac{f_k}{k+1}}
\left[t+\frac{\vec{r}}{c}\cdot\hat{n}\right]^{k+1}\right)
\end{equation}

where $\vec{r}(t)$ denotes the detector position,
$\hat{n}=(\sin{\theta}\cos{\phi},\sin{\theta}\sin{\phi},\cos{\theta})$
is a unit vector in the direction of the source. The angles
$\theta$, $\phi$ are associated with the pulsar source;
$\vec{v}(t)=\frac{d\vec{r}(t)}{dt}$ is the velocity of the
detector and  $f_k$ are the spin down parameters. Here $|f_k|\leq
\tau_{min}^{-k}$ and $\tau_{min}$ is the spin down age in years.
Here $\tau=\frac{f}{df/dt}$ is life in years of the pulsars. For
the extreme case of the gravitational-wave frequency of $10^3$ Hz,
the spin-down age is $\tau=40$ years for pulsars in our galaxy and
$\geq$ $10^7$ years for millisecond pulsars. Our analysis
considers the equality situation, that is,
$|f_k|=\tau_{min}^{-k}$. The phase $\phi$ depends on the frequency
$f_0$, $k$ spin-down parameters, and on the angles $\phi$,
$\theta$ and the co-latitude $\alpha$ of the detector.

In the earlier analysis of Valluri et.al. \cite{CQG2002}, the spin
down corrections were neglected. In their analysis of the FM of a
monochromatic plane wave, they characterize the motion of the
Earth (and detector) by: (a) assuming the orbit of the Earth to be
circular. (b) neglecting the effect of the Moon and the
perturbation effects of Jupiter on the Earth's orbit. In this
work, in addition to incorporating spin down effects, we also
include the approximate correction for the eccentricity of the
orbital motion of the Earth.

With these assumptions, (a) and (b), and ignoring spin down
corrections we arrive at the following expression of the FT for
the GW pulsar signal \cite{CQG2002}, rewritten more concisely as,

\begin{equation}
\widetilde{h}(f)=S_{n l m}(\omega _{0},\omega
_{orb},T_{Er},n,l,m,A,R,k,\alpha ,\theta ,\phi)=\underset{n =
-\infty }{\overset{\infty }{\sum }} \underset{l =
0}{\overset{\infty }{\sum }}\overset{l }{\underset{m = -l }{\sum
}} \psi_0 \psi_1 \psi_2 \psi_3 \psi_4
\end{equation}

where

\begin{equation}
\psi_0(n,l,m,\alpha,\theta,\phi)= 4\pi i^{l}Y_{l m}(\theta ,\phi
)N_{l m}P_{l }^{m}(\cos \alpha )
\end{equation}

\begin{equation}
\psi_1(n,\theta, \phi, T_{Er}, f_0,
A)=T_{Er}\sqrt{\frac{\pi}{2}}e^{-i\frac{2\pi f_{0}A}{c}\sin \theta
\cos \phi } i^{n}e^{-in\phi }J_{n}\left( \frac{2\pi f_{0}A\sin
\theta }{c}\right)
\end{equation}

\begin{equation}
\psi_2(l,\omega_{orb}, \omega_{r}, n, m, R)=\left\{\frac{1-e^{i\pi
(l -B_{orb})R}}{1-e^{i\pi (l -B_{orb})}} \right\}
2e^{-iB_{orb}\frac{\pi}{2}}\frac{1}{2^{2l +1}}
\end{equation}

\begin{equation}
\psi_3(k,l,m,n,\omega_{orb}, \omega_r)=k^{l
+\frac{1}{2}}\frac{\Gamma \left(l +1\right) }{\Gamma \left(l
+\frac{3}{2}\right)\Gamma \left(\frac{l
+B_{orb}+2}{2}\right)\Gamma \left(\frac{l -B_{orb}+2}{2}\right)}
\end{equation}

\begin{equation}
\psi_4(k,l,m,n,\omega_{orb}, \omega_r)=_1F_{3}\left(l +1;l
+\frac{3}{2},\frac{l +B_{orb}+2}{2},\frac{l
-B_{orb}+2}{2};\frac{-k^{2}}{16} \right)
\end{equation}\\

The angle $\alpha$ is the co-latitude detector angle and angles
$\theta$, $\phi$ are associated with the pulsar source. Here
$\omega_0=2\pi f_0$, $\omega_{orb}=\frac{2\pi}{T_{orb}}$
($T_{orb}=365$ days, $T_{Er} = 1$ day),
$B_{orb}=2\left(\frac{\omega-\omega_0}{\omega_r}+\frac{m}{2}+\frac{n
\omega_{orb}}{\omega_{rot}}\right)$, $k=\frac{4\pi f_0 R_E
\sin(\alpha)}{c}$ ($R_E$ is the radius of Earth, $c$ is the
velocity of light) and $A$ is the sun-earth distance.

\section{Calculation of the FT including Spin Down Corrections}

Considering the spin down corrections in Equation (2), we have for
the summation term in $\phi(t;\lambda)$ in Equation (3),
\begin{center}
$\overset{\infty}{\underset{k=0}{\sum}}\frac{f_{k}}{k+1}\left[t+\frac{\vec{r}\cdot\hat{n}}{c}\right]
^{k+1}$\\
\end{center}

Upon using the definition, $|f_k|\equiv\tau_{min}^{-k}$ we have
for this summation,
\begin{center}
$\overset{\infty}{\underset{k=0}{\sum}}\frac{\tau_{min}}{\tau_{min}^{k+1}}
\frac{1}{k+1}\left[t+\frac{\vec{r}\cdot\hat{n}}{c}\right]
^{k+1}$\\
\end{center}

This power series represents a logarithmic function of the form
$-\ln(1-x)$. It is interesting to note that this is the $n=1$ case
of the multi-valued polylogarithm function $L_{n}(x)$ and the
phase evolution, apart from the factor $i \pi f_0$ is one of the
versions of the multi-valued Lambert-W function \cite{MAX2003}.
Hence, we have,
\begin{center}
$-\tau_{min}\ln \left( 1-\frac{t+\frac{\vec{r}\cdot
\hat{n}}{c}}{\tau_{min}}\right)$\\

$\therefore \phi (t) = $exp$\left[i\pi f_{0}\left
\{(t+\frac{\vec{r}\cdot \hat
{n}}{c})-\tau_{min}\ln(1-\frac{t+\frac{\vec{r}\cdot
\hat{n}}{c}}{\tau_{min}})\right\}\right]$\\

$= e^{^{i\pi f_{0}(t+\frac{\vec{r}\cdot\hat{n}}{c})}} \exp
\left[-i\tau_{min}\pi f_{0} \ln
(1-\frac{t+\frac{\vec{r}\cdot\hat{n}}{c}}{\tau_{min}})\right]$

\end{center}

\begin{equation}
= e^{^{i\pi f_{0}(t+\frac{\vec{r}\cdot \hat{n}}{c})}}
\left[1-\frac{t+\frac{\vec{r}\cdot\hat{n}}{c}}{\tau_{min}}\right]^{-i\tau_{min}\pi
f_{0}}
\end{equation}

Using the binomial expansion on equation (10), we obtain,
\begin{equation}
e^{^{i\pi f_{0}(t+\frac{\vec{r}\cdot \hat{n}}{c})}} \left[1+i\pi
f_{0}\tau_{min}\frac{t+\frac{\vec{r}\cdot\hat{n}}{c}}{\tau_{min}}
+ \frac{(i\pi f_0 \tau_{min})(i\pi f_0
\tau_{min}+1)}{2}\left(\frac{t+\frac{\vec{r}\cdot\hat{n}}{c}}{\tau_{min}}\right)^2
+...\right]
\end{equation}

We define the Spindown Moment Integrals as follows:

\begin{equation}
\int_{0}^{T}e^{i\pi f_{0}\left(t+\frac{\vec{r}\cdot
\hat{n}}{c}\right)} \cdot e^{-i2\pi ft}\cdot \left(
t+\frac{\vec{r}\cdot\hat{n}}{c}\right)^{k}dt
\end{equation}

for $k=$ 0,1,2,3,...., where $k=0$ denotes the absence of spin
down corrections and recovers the formula for the FT given in
Equations (4-9).

Thus, a generic spin down moment integral could be written from
the derivative of a generic integral,

\begin{equation}
I_{generic}=\int_{0}^{T} e^{\left(i\pi
f_{0}(t+\frac{\vec{r}\cdot\hat{n}} {c})- i 2\pi f t \right)}dt
\end{equation}

Hence, we have the $k-$th derivative as,
\begin{equation}
\frac{d^{k}I_{generic}}{df_{0}^{k}}= \int_{0}^{T}\left[i\pi
\left(t+ \frac{\vec{r}\cdot\hat{n}}{c}\right)\right]^{k} e^{[i\pi
f_{0}(t+\frac{\vec{r}\cdot \hat{n}}{c})-i2\pi ft]} dt
\end{equation}

for $k=$0,1,2,3,......\\

This $k$ derivative gives the Corresponding Spin down moment
integral. Thus, $I_{generic}$ has been analytically evaluated. Its
derivatives with respect to $f_{0}$ can be done by symbolic
packages like Maple, Mathematica etc. It should be noted that the
FT of the GW signal using the later parametrization given by Brady
and Creighton (2000) can also be analytically incorporated in a
similar manner of derivation given above.

It should be noted that previously these spin down corrections
were incorporated by splitting up a time interval, say $T_0$ into
$M$ equal parts each of interval $\Delta t$ $(T_0=M\Delta t)$ so
that the signal is monochromatic in each interval or "window"
\cite{SS2002}. This is the idea behind "stacking" where FTs are
incoherently combined by adding their power spectra. "Tracking" as
suggested by Brady et. al (2000), Papa and Schutz, (cited in
review article gr-qc/9802020) is the attempt to track weak lines
in individual FTs in successive data sets to identify persistent
signals. We express our final result as a coherent sum over any
given time interval as a closed form solution.\\

In our original FT, we have not included the effects of orbital
eccentricity which arise due to the orbital motion of Earth.
Denoting the orbital eccentricity by $e_{\bigoplus}(=0.017)$, we
find that by replacing the usual Sun-Earth distance, $A$ by the
following modified $A$, is a reasonable estimate of the
correction.

\begin{equation}
A \rightarrow A\sqrt{1-e_{\bigoplus}^2}
\end{equation}

The correction in the phase of the GW signal due to the effects of
orbital eccentricity can also be accurately estimated as has
already been pointed out by Jaranowski et al. (1998).

\section{Conclusions}

Recently, new mechanisms e.g. r-mode instability of spinning
neutron stars and temperature asymmetry in the interior of the
neutron star with misaligned spin axis have been discussed in the
literature \cite{LIGOR,SS2002}. The continuous GW signal may
consist of frequencies which are multiple of some basic
frequencies. Brady and Creighton have given estimates of the
characteristic strain of gravitational waves from an active r-mode
instability in a newborn neutron star suggesting that these
sources will be detectable by the enhanced interferometers in LIGO
out to distances $~8$ Mpc; the rate of supernovae is $~0.6$ per
year within this distance \cite{JKS1998}, \cite{BC2000}.

We have presented in this paper the rudiments of a simple analysis
for spin down for sources of continuous gravitational waves. For
the more computationally-intensive search over all sky positions
and spin down parameters, it is important to be able to calculate
the smallest number of independent parameter values which must be
sampled in order to cover the entire space of signals. The PWESH
has the potential to improve the numerical accuracy and
convergence of analytic FT's and spin down corrections associated
with the GW signal.

The study of templates in time-domain has been made by many
research workers(Schutz 1991; Krolak 1997; Brady et al. 1998;
Brady, Creighton 2000; Jaranowski et al. 1998; Jaranowski, Krolak
1999, 2000 \cite{JK1999}). However, the analysis in the frequency
domain has the advantage of incorporating the spectral noise
density of the interferometer. The data output at the
interferometer is available in discrete form and the question
arises if the analytical FT is in fact a very convenient tool.
However, the FFT has a resolution limited to $1/T_0$ as pointed by
Srivastava and Sahay \cite{SS2002}. It is hoped that the
analytical FT will provide useful insights.

We intend to further estimate from our spin down analysis the
number of templates required for matched filtering analysis. We
intend to make use of our approach developed for all frequencies
to evaluate the Fitting Factor (FF) as given by Apostolatos
\cite{AP} and considered for low frequencies by Srivastava and
Sahay (2002). The time-delay type effects \cite{BACK} due to
general relativity and other corrections for the ephemeris have
been treated in detail by Cutler in his Ligo Algorithmic Library
(LAL) Barycenter Package (available at:
http://www.lsc-group.phys.uwm.edu/) and will also be taken into
account.

A new expansion was explored by MacPhie et.al. \cite{MWIEEE} both
in the case of scalar and vector harmonics. This can be extended
to tensor harmonics as well, and work in that direction will be
carried out. This would be of possible relevance to study the
multipole formalisms for gravitational radiation \cite{KThorne80}.

\section{Acknowledgments}
We are deeply grateful to SHARCNET (Shared Hierarchical Academic
Research Cluster Network) for valuable grant supplement to Adam
Vajda that made this study feasible. We are also greatly indebted
to Drs. Nico Temme (CWI, Amsterdam), Alessandra Maria Papa (AEI,
Potsdam), Tom Prince (JPL, Pasadena) and B. S. Sathyaprakash (U.
Cardiff)for valuable suggestions.

\end{document}